\documentclass[12pt]{iopart}

\newcommand{\eqref}[1]{(\ref{#1})}

\usepackage{iopams}
\usepackage{graphicx}
\usepackage{amsfonts}
\usepackage{amsbsy}
\usepackage{amssymb}
\usepackage{color}

\newcommand{\beq}{\begin{equation}}
\newcommand{\eeq}{\end{equation}}
\newcommand{\beqa}{\begin{eqnarray}}
\newcommand{\eeqa}{\end{eqnarray}}
\newcommand{\bem}{\begin{multline}}
\newcommand{\eem}{\end{multline}}

\begin{document}

\title[On anomalous diffusion and the out of equilibrium response function...]{On anomalous diffusion and the out of equilibrium response function in one-dimensional models}

\author{D Villamaina, A Sarracino, G Gradenigo, A Puglisi and A Vulpiani}
\address{CNR-ISC and Dipartimento di Fisica, Universit\`a Sapienza - p.le A. Moro 2, 00185, Roma, Italy}

\ead{dario.villamaina@roma1.infn.it,alessandro.sarracino@roma1.infn.it\\
ggradenigo@gmail.com,andrea.puglisi@roma1.infn.it\\
angelo.vulpiani@roma1.infn.it}


\begin{abstract} We study how the Einstein relation
    between spontaneous fluctuations and the response to an external
    perturbation holds in the absence of currents, for the comb model
    and the elastic single-file, which are examples of systems with
    subdiffusive transport properties.  The relevance of
    nonequilibrium conditions is investigated: when a stationary
    current (in the form of a drift or an energy flux) is present, the
    Einstein relation breaks down, as it is known to happen in systems
    with standard diffusion.  In the case of the comb model, a general
    relation - appeared in the recent literature - between response
    function and an unperturbed suitable correlation function, allows
    us to explain the observed results.  This suggests that a relevant
    ingredient in breaking the Einstein formula, for stationary
    regimes, is not the anomalous diffusion but the presence of
    currents driving the system out of equilibrium.
\end{abstract}

\maketitle

\section{Introduction}

In his seminal paper on the Brownian Motion, Einstein, beyond 
the celebrated relation between the diffusion coefficient $D$ and the
Avogadro number, found the first example of fluctuation-dissipation
relation (FDR).
In the absence of external forcing one has, for large times $t\to\infty$,
\begin{equation}
\label{1}
\langle x(t) \rangle=0 \,\,\,\,\, ,  \,\,\,\,\,
\langle x^2(t) \rangle \simeq 2 D t \,\, ,
\end{equation}
where $x$ is the position of the Brownian particle and
the average is taken over the unperturbed dynamic. Once a small
constant external force $F$ is applied one has a linear drift
\begin{equation}
\label{2}
\overline{\delta x}(t)=
\langle x(t) \rangle_F - \langle x(t) \rangle
 \simeq \mu F t \,\,
\end{equation}
where $\langle\ldots\rangle_F$ indicates the average on the
perturbed system, and $\mu$ is the mobility of the colloidal particle.
 It is  remarkable that $\langle x^2(t) \rangle$
is proportional to $\overline{\delta x}(t)$ at any time:
\begin{equation}
\label{3}
{\langle x^2(t) \rangle  \over \overline{\delta x}(t)}=\frac{2}{\beta F},
\end{equation}
and the Einstein relation (a special case of the
fluctuation-dissipation theorem~\cite{K66}) holds: $\mu = \beta D$,
with $\beta=1/k_BT$ the inverse temperature and $k_B$ the Boltzmann constant.

On the other hand it is now well established that beyond the standard
diffusion, as in~(\ref{1}), one can have systems with anomalous
diffusion (see for instance~\cite{BG90,GSGWS96,CMMV99,MK00,BC05}), i.e.
\begin{equation}
\label{4}
\langle x^2(t)\rangle \sim t^{2 \nu} \,\,\, \mbox{with} \,\,\,  \nu \ne
1/2.
\end{equation}
Formally this corresponds to have $D=\infty$ if $\nu > 1/2$
(superdiffusion) and $D=0$ if $\nu < 1/2$ (subdiffusion).  In this
letter we will limit the study to the case $\nu < 1/2$.  It is quite
natural to wonder if (and how) the FDR
changes in the presence of anomalous diffusion, i.e. if instead
of~(\ref{1}), Eq.~(\ref{4}) holds.  In some systems it has been showed
that~(\ref{3}) holds even in the subdiffusive case.  This has been
explicitly proved in systems described by a fractional-Fokker-Planck
equation~\cite{MBK99}, see also~\cite{BF98,CK09}. In addition there
is clear analytical~\cite{LATBL10} and numerical~\cite{VPV08}
evidences that~(\ref{3}) is valid for the elastic single file, i.e. a
gas of hard rods on a ring with elastic collisions, driven by an
external thermostat, which exhibits subdiffusive behavior, $\langle
x^2 \rangle \sim t^{1/2}$~\cite{HKK96}.

The aim of this paper is to discuss the validity of the
fluctuation-dissipation relation in the form~(\ref{3}) for systems with
anomalous diffusion which are not fully described by a
fractional Fokker-Planck equation.  In particular we will investigate
the relevance of the anomalous diffusion, the presence of non
equilibrium conditions and the (possible) role of finite size.
Since we are also interested in the study of transient
regimes, we will consider models with microscopic dynamics 
described in terms of transition rates or microscopic
interactions.

First, we focus on the study of a particle moving on a
``finite comb'' lattice with teeth of size $L$~\cite{R01}.  In the
limit $L=\infty$ an anomalous subdiffusive behavior, $\langle x^2
\rangle \sim t^{1/2}$, holds and the system can be mapped, for large
times, onto a continuous time random walk~\cite{R01}.  For finite $L$
the subdiffusion is only transient and at very large time $t>t^*(L)
\sim L^2$ one has a standard diffusion: $\langle x^2 \rangle \sim
t$. We will see that Eq.~(\ref{3}), where in this case
the perturbed average is obtained with unbalanced transition rates
driving the particle along the backbone of the comb, holds both for
$t>t^*(L)$ and $t<t^*(L)$ with the same constant.  This in spite of
the fact that the probability densities $P(x,t)$ in the two regimes
are very different.  The scenario changes in the presence of ``non
equilibrium'' conditions, i.e. with a drift, which induces a current,
in the unperturbed state: the relation~(\ref{3}) does not hold
anymore.  On the other hand, in this case it is possible
to use a generalized fluctuation-dissipation relation, derived by
Lippiello \emph{et al.} in~\cite{LCZ05}, which gives the response
function in terms of unperturbed correlation functions and is an
example of non equilibrium FDR valid under rather general
conditions~\cite{CKP94,D05,LCZ05,ss06,BPRV08,BMW09,SS10,CLSZ10}.  A
generalization of the Einstein formula was also proved in the
framework of continuous time random walks in~\cite{HBMB08}.  So we can
say that the Einstein relation~(\ref{3}) also holds in 
cases with anomalous diffusion when no current is present, but it is
necessary to introduce suitable corrections when a perturbation is applied
to a system with non zero drift.

In addition we compare the results found in comb models, with those
obtained for single-file diffusion with a finite number of particles.
There we will also consider a non equilibrium case, with
the introduction of inelastic collisions which induce an energy flux
crossing the system. Our results suggest that the presence of non
equilibrium currents plays a relevant role in modifying Eq.~\eqref{3}
in stationary states.\\

\section{Comb: diffusion and response function}

The comb lattice is a discrete structure consisting of an infinite
linear chain (backbone), the sites of which are connected with other
linear chains (teeth) of length $L$~\cite{R01}. We denote
by $x\in(-\infty,\infty)$ the position of the particle performing the
random walk along the backbone and with $y\in[-L,L]$ that along a
tooth. The transition probabilities from $(x,y)$ to $(x',y')$ are:
\begin{eqnarray}
W^d[(x,0)\rightarrow (x\pm 1,0)]&=&1/4\pm d \nonumber \\
W^d[(x,0)\rightarrow (x,\pm 1)]&=&1/4 \nonumber \\
W^d[(x,y)\rightarrow (x,y\pm 1)]&=&1/2~~~ \textrm{for}~y\ne 0,\pm L.
\label{ww}
\end{eqnarray}
On the boundaries of each tooth, $y=\pm L$, the particle is reflected
with probability 1. The case $L=\infty$ is obtained in
numerical simulations by letting the $y$ coordinate increase without
boundaries. Here we consider a discrete time process and, of course,
the normalization $\sum_{(x',y')}W^d[(x,y)\rightarrow (x',y')]=1$
holds.  The parameter $d\in[0,1/4]$ allows us to consider also the
case where a constant external field is applied along the $x$ axis,
producing a non zero drift of the particle. A state with a non zero
drift can be considered as a perturbed state (in that case we denote
the perturbing field by $\varepsilon$), or it can be itself the
starting state where a further perturbation can be added changing
$d\rightarrow d+\varepsilon$.

\begin{figure}
\includegraphics[scale=0.325,clip=true]{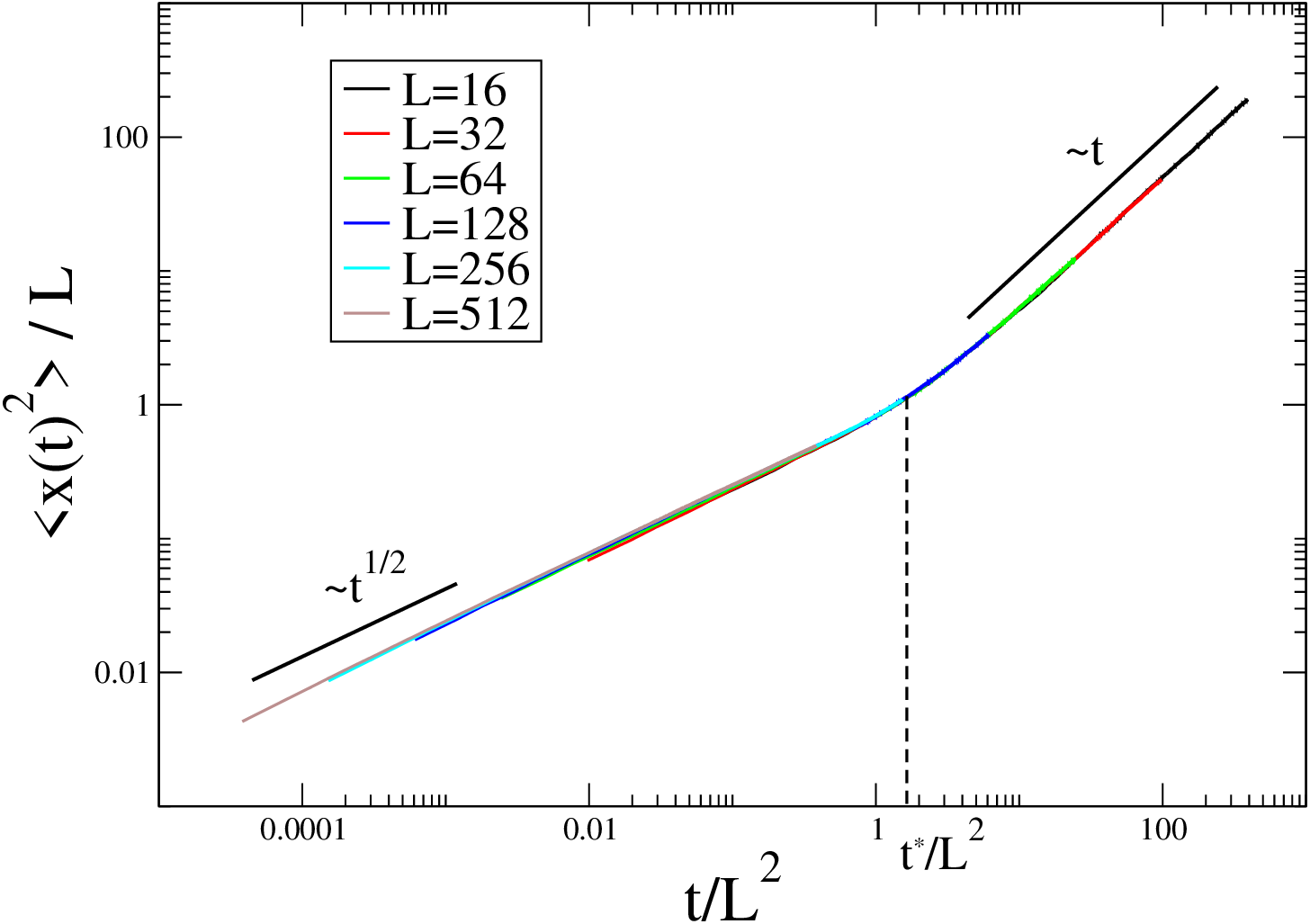}
\includegraphics[scale=0.32,clip=true]{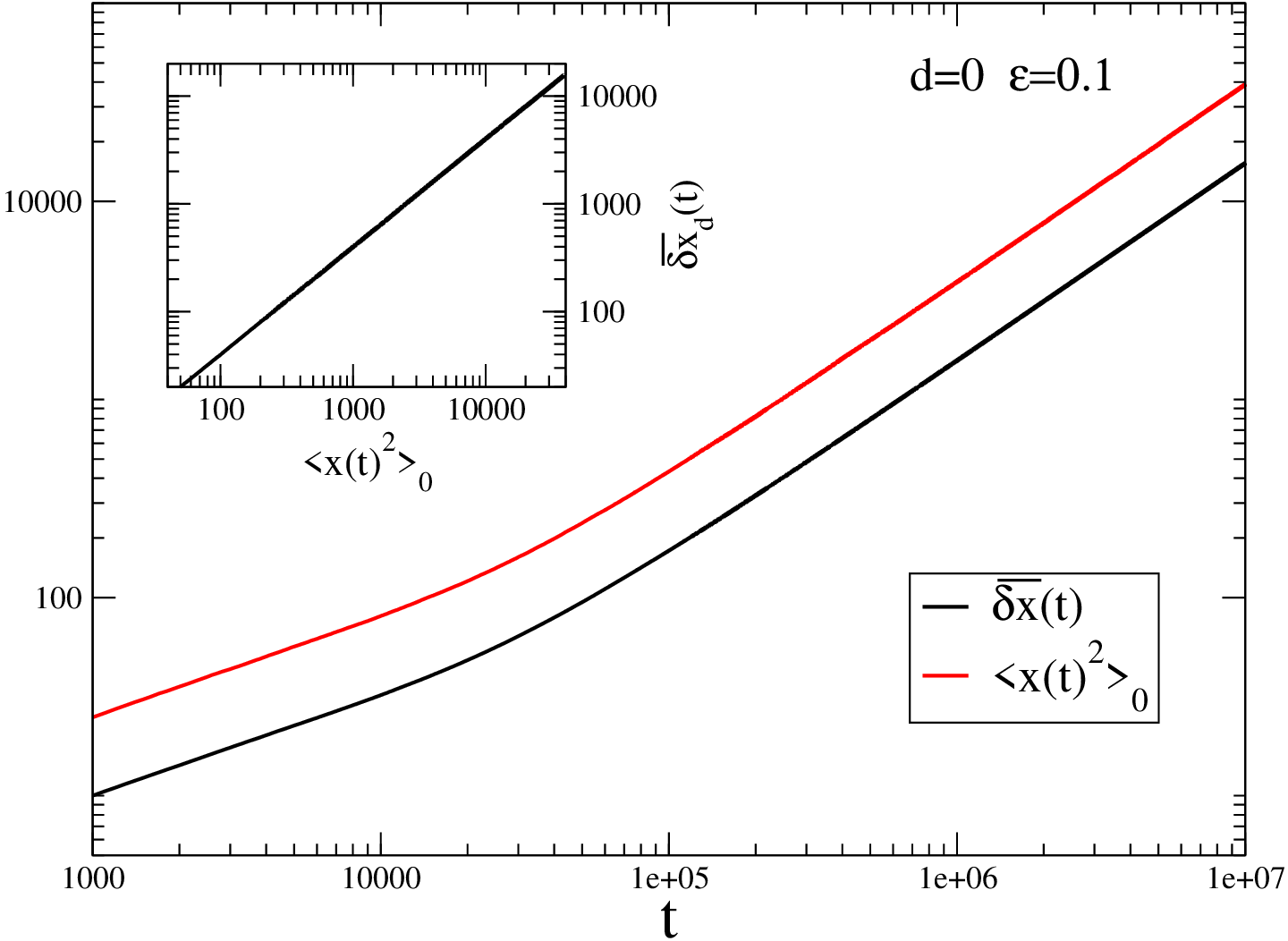}
\caption{Left panel: $\langle x^2(t) \rangle_0/L$ \emph{vs}
$t/L^2$ is plotted for several values of $L$ in the comb model. Right
panel: $\langle x^2(t) \rangle_0$ and the response function
$\overline{\delta x}(t)$ for $L=512$. In the inset the
parametric plot $\overline{\delta x}(t)$ \emph{vs} $\langle x^2(t)
\rangle_0$ is shown.}
\label{fig1}
\end{figure}

Let us start by considering the case $d=0$.  For finite teeth length
$L<\infty$, we have numerical evidence of a dynamical
crossover from a subdiffusive to a simple diffusive asymptotic
behaviour (see Fig.~\ref{fig1})
\begin{equation} 
\langle x^2(t) \rangle_0 \simeq
\left\{\begin{array}{cc}
C t^{1/2} &  t < t^{*}(L)\nonumber
\\ 
2 D(L) t &  t >t^{*}(L),
\label{nodrift2}
\end{array}\right.
\end{equation} 
where $C$ is a constant and $D(L)$ is an effective diffusion
coefficient depending on $L$. The symbol $\langle\ldots\rangle_0$
denotes an average over different realizations of the
dynamics~(\ref{ww}) with $d=0$ and initial condition $x(0)=y(0)=0$.
We find $t^*(L)\sim L^2$ and $D(L)\sim 1/L$ and in the left panel of
Fig.~\ref{fig1} we plot $\langle x^2(t) \rangle_0/L$ as function of
$t/L^2$ for several values of $L$, showing an excellent data collapse.

In the limit of infinite teeth, $L \rightarrow \infty$,
$D\rightarrow 0$ and $t^*\rightarrow\infty$ and the system shows
a pure subdiffusive behaviour~\cite{HBA97} 
\beq \langle x^2(t) \rangle_0
\sim t^{1/2}.
\label{nodrift1}
\eeq 
In this case, the probability distribution function behaves as
\begin{equation}
P_0(x,t)\sim t^{-1/4}e^{-c\left(\frac{|x|}{t^{1/4}}\right)^{4/3}},
\label{4/3}
\end{equation}
where $c$ is a constant, in agreement with an argument \emph{\`a la}
Flory~\cite{BG90}. The behaviour~(\ref{4/3}) also holds in the case of
finite $L$, provided that $t<t^*$. For larger times a simple Gaussian
distribution is observed. Note that, in general, the
scaling exponent $\nu$, in this case $\nu=1/4$, does not determine
univocally the shape of the pdf.  Indeed, for the single-file model,
discussed below, we have the same $\nu$ but the pdf is
Gaussian~\cite{WBL00}.

In the comb model with infinite teeth, the FDR in its standard form is
fulfilled, namely if we apply a constant perturbation $\varepsilon$
pulling the particles along the 1-d lattice one has numerical evidence that
\beq 
\langle x^2(t)\rangle_0 \simeq C\overline{\delta x}(t)\sim t^{1/2}.
\label{FDRst}
\eeq 
In the following section we derive this result from a generalized
FDR.  Moreover, the proportionality between $\langle x^2(t)\rangle_0$
and $\overline{\delta x}(t)$ is fulfilled also with $L<\infty$, where
both the mean square displacement (m.s.d.) and the drift with an
applied force exhibit the same crossover from subdiffusive, $\sim
t^{1/2}$, to diffusive, $\sim t$ (see Fig.~\ref{fig1}, right panel).
Therefore what we can say is that the FDR is somehow ``blind" to the
dynamical crossover experienced by the system.  When the perturbation
is applied to a state without any current, the proportionality between
response and correlation holds despite anomalous transport phenomena.

Our aim here is to show that, differently from what depicted above
about the zero current situation, within a state with a non zero
drift~\cite{BCGR03} the emergence of a dynamical crossover is
connected to the breaking of the FDR.  Indeed, the m.s.d. in the
presence of a non zero current, even with $L=\infty$, shows a dynamical
crossover
\begin{equation}
\langle x^2(t) \rangle_d \sim  a~t^{1/2} + b~t,
\label{drift1}
\end{equation}
where $a$ and $b$ are two constants, whereas
\begin{equation} 
\overline{\delta x}_d(t) \sim t^{1/2},
\label{drift2}
\end{equation}
with $\overline{\delta x}_d(t)=\langle
x(t)\rangle_{d+\varepsilon}-\langle x(t)\rangle_d$: at large times
the Einstein relation breaks down (see Fig.~\ref{figFDR}).  The
proportionality between response and fluctuations cannot be recovered
by simply replacing $\langle x^2(t) \rangle_d$ with $\langle x^2(t)
\rangle_d-\langle x(t)\rangle^2_d$, as it happens for Gaussian
processes (see discussion below), namely we find numerically 
\begin{equation}
\langle [x(t)-\langle x(t)\rangle_d]^2 \rangle_d \sim a'~t^{1/2}+b'~t,
\label{deltax2comb}
\end{equation}
where $a'$ and $b'$ are two constants, as reported in Fig.~\ref{figFDR}.

\section{Comb: application of a generalized FDR}

The discussion of the previous section shows that the first moment of
the probability distribution function with drift $P_d(x,t)$ and the
second moment of $P_0(x,t)$ are always proportional.  
Note that in the presence of a drift the pdf is strongly asymmetric
with respect to the mean value, as shown in Fig.~\ref{pdf} for a
system with $L=\infty$.  Differently, the first moment of
$P_{d+\varepsilon}(x,t)$ is not proportional to the second moment of
$P_{d}(x,t)$, namely $\langle x(t)\rangle_{d+\varepsilon}\nsim \langle
x^2(t)\rangle_d-\langle x(t)\rangle^2_d$.  In order to find out a
relation between such quantities, we need a generalized
fluctuation-dissipation relation.

According to the definition~(\ref{ww}), one has for the backbone
\begin{eqnarray}
\hspace*{-2.5cm}W^{d+\varepsilon}[(x,y)\rightarrow (x',y')]=W^d[(x,y)\rightarrow (x',y')]
\left(1+\frac{\varepsilon (x'-x)}{W^0+d (x'-x)}\right)
\simeq W^de^{\frac{\varepsilon}{W^0}(x'-x)}, \nonumber \\
\end{eqnarray}
where $W^0=1/4$, and the last
expression holds under the condition $d/W^0\ll 1$.
Regarding the above expression as a \emph{local detailed balance}
condition for our Markov process we can rewrite it, for
$(x,y)\ne (x',y')$, as
\begin{equation}
W^{d+\varepsilon}[(x,y)\rightarrow (x',y')]=W^d[(x,y)\rightarrow (x',y')] 
e^{\frac{h(\varepsilon)}{2}(x'-x)},
\label{Wpert}
\end{equation}
where $h(\varepsilon)=2\varepsilon/W^0$.  For general
models where the perturbation enters the transition probabilities
according to Eq.~(\ref{Wpert}), the following formula for the
integrated linear response function has been
derived~\cite{LCZ05,BMW09,CLSZ10}
\begin{equation}
\hspace*{-2cm}\frac{\overline{\delta\mathcal{O}}_d}{h(\varepsilon)}=\frac{\langle \mathcal{O}(t)\rangle_{d+\varepsilon}- \langle
\mathcal{O}(t)\rangle_d}{h(\varepsilon)}
=\frac{1}{2}\left[\langle\mathcal{O}(t)x(t)\rangle_d-\langle\mathcal{O}(t)x(0)\rangle_d
-\langle\mathcal{O}(t)A(t,0)\rangle_d\right],
\label{FDR}
\end{equation}
where $\mathcal{O}$ is a generic observable, and $A(t,0)=\sum_{t'=0}^t
B(t')$, with
\begin{equation}
B[(x,y)]=\sum_{(x',y')}(x'-x)W^d[(x,y)\rightarrow (x',y')].
\end{equation}
The above observable yields an effective measure of the propensity of
the system to leave a certain state $(x,y)$ and, in some contexts, it
is referred to as \emph{activity}~\cite{ARDLW08}.  Recalling the
definitions~(\ref{ww}), from the above equation we have
$B[(x,y)]=2d\delta_{y,0}$ and therefore the sum on $B$ has an
intuitive meaning: it counts the time spent by the particle on the $x$
axis.  The results described in the previous section can be then read
in the light of the fluctuation-dissipation relation~(\ref{FDR}):

\begin{figure}[t!]
\centering
\includegraphics[scale=0.4,clip=true]{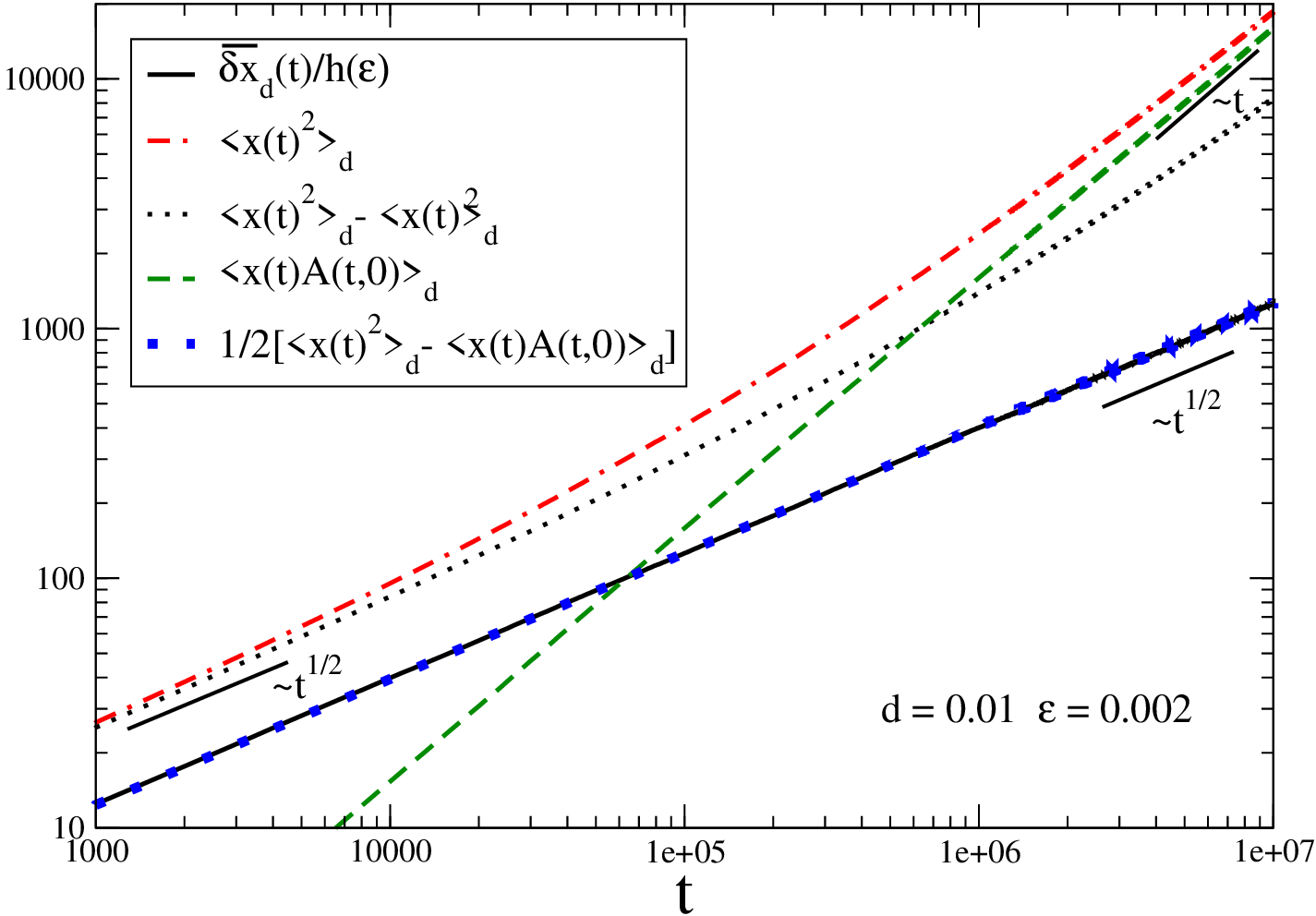}
\caption{Response function (black line), m.s.d. (red
dotted line) and second cumulant (black dotted line) measured in the
the comb model with $L=\infty$, field $d=0.01$ and perturbation
$\varepsilon=0.002$. The correlation with activity (green dotted
line) yields the right correction to recover the full response
function (blue dotted line), in agreement with the FDR~(\ref{FDR}).}
\label{figFDR}
\end{figure}

i) Putting $\mathcal{O}(t)=x(t)$, in the case without drift, i.e. $d=0$, one has $B=0$ and,
recalling the choice of the initial condition $x(0)=0$, 
\begin{equation}
\frac{\overline{\delta x}}{h(\varepsilon)}=\frac{\langle x(t)\rangle_{\varepsilon}- \langle x(t)\rangle_0}{h(\varepsilon)}=
\frac{1}{2}\langle x^2(t)\rangle_0.
\label{FDR1}
\end{equation}
This explains the observed behaviour (\ref{FDRst}) even in the
anomalous regime and predicts the correct proportionality factor,
$\overline{\delta x}(t)=\varepsilon/W^0\langle x^2(t)\rangle_0$.

ii) Putting $\mathcal{O}(t)=x(t)$, in the case with $d\ne 0$, one has
\begin{equation}
\frac{\overline{\delta x}_d}{h(\varepsilon)}
=\frac{1}{2}\left[\langle x^2(t)\rangle_d-\langle x(t)A(t,0)\rangle_d\right].
\end{equation}
This explains the observed behaviours~(\ref{drift1})
and~(\ref{drift2}): the leading behavior at large times of $\langle
x^2(t)\rangle_d \sim t$, turns out to be exactly canceled by the term
$\langle x(t)A(t,0)\rangle_d$, so that the relation between response
and unperturbed correlation functions is recovered (see
Fig.~\ref{figFDR}).

iii) As discussed above, it is not enough to substitute $\langle
x^2(t)\rangle_d$ with $\langle x^2(t)\rangle_d-\langle
x(t)\rangle_d^2$ to recover the proportionality with $\overline{\delta
x}_d(t)$ when the process is not Gaussian. This can be explained in
the following manner. By making use of the second order out of equilibrium FDR derived
by Lippiello \emph{et al.} in~\cite{LCSZ08a,LCSZ08b,CLSZ10b}, which is needed due to the
vanishing of the first order term for symmetry, we can explicitly
evaluate
\begin{equation}
\langle x^2(t)\rangle_d
=\langle x^2(t)\rangle_0+h^2(d)\frac{1}{2}\left[\frac{1}{4}\langle x^4(t)\rangle_0
+\frac{1}{4}\langle x^2(t)A^{(2)}(t,0)\rangle_0\right],
\label{FDR2}
\end{equation}
where $A^{(2)}(t,0)=\sum_{t'=0}^t B^{(2)}(t')$ with $B^{(2)}=-\sum_{x'}(x'-x)^2W[(x,y)\rightarrow (x',y')]=-1/2\delta_{y,0}$.
Then, recalling Eq.~(\ref{FDR1}), we obtain
\begin{equation}
\hspace*{-2.5cm}\langle x^2(t)\rangle_d- \langle
x(t)\rangle^2_d= \langle
x^2(t)\rangle_0+h^2(d)\left[\frac{1}{8}\langle x^4(t)\rangle_0
+\frac{1}{8}\langle x^2(t)A^{(2)}(t,0)\rangle_0-\frac{1}{4}\langle
x^2(t)\rangle^2_0\right].
\label{FDR3}
\end{equation}
Numerical simulations show that the term in the square brackets grows
like $t$ yielding a scaling behaviour with time consistent with
Eq.~(\ref{deltax2comb}).  On the other hand, in the case of the simple
random walk, one has $B^{(2)}=-1$ and $A^{(2)}(t,0)=-t$ and then
\begin{equation}
\hspace*{-1cm}\langle x^2(t)\rangle_d- \langle
x(t)\rangle^2_d= \langle
x^2(t)\rangle_0+h^2(d)\left[\frac{1}{8}\langle x^4(t)\rangle_0
-\frac{1}{8}t\langle x^2(t)\rangle_0-\frac{1}{4}\langle
x^2(t)\rangle^2_0\right].
\label{FDR4}
\end{equation}
Since in the Gaussian case $\langle x^4(t)\rangle_0=3\langle
x^2(t)\rangle^2_0$ and $\langle x^2(t)\rangle_0=t$, the term in the
square brackets vanishes identically and that explains why, in the
presence of a drift, the second cumulant grows \emph{exactly} as the
second moment with no drift.

\begin{figure}[ht!]
\centering
\includegraphics[scale=0.4,clip=true]{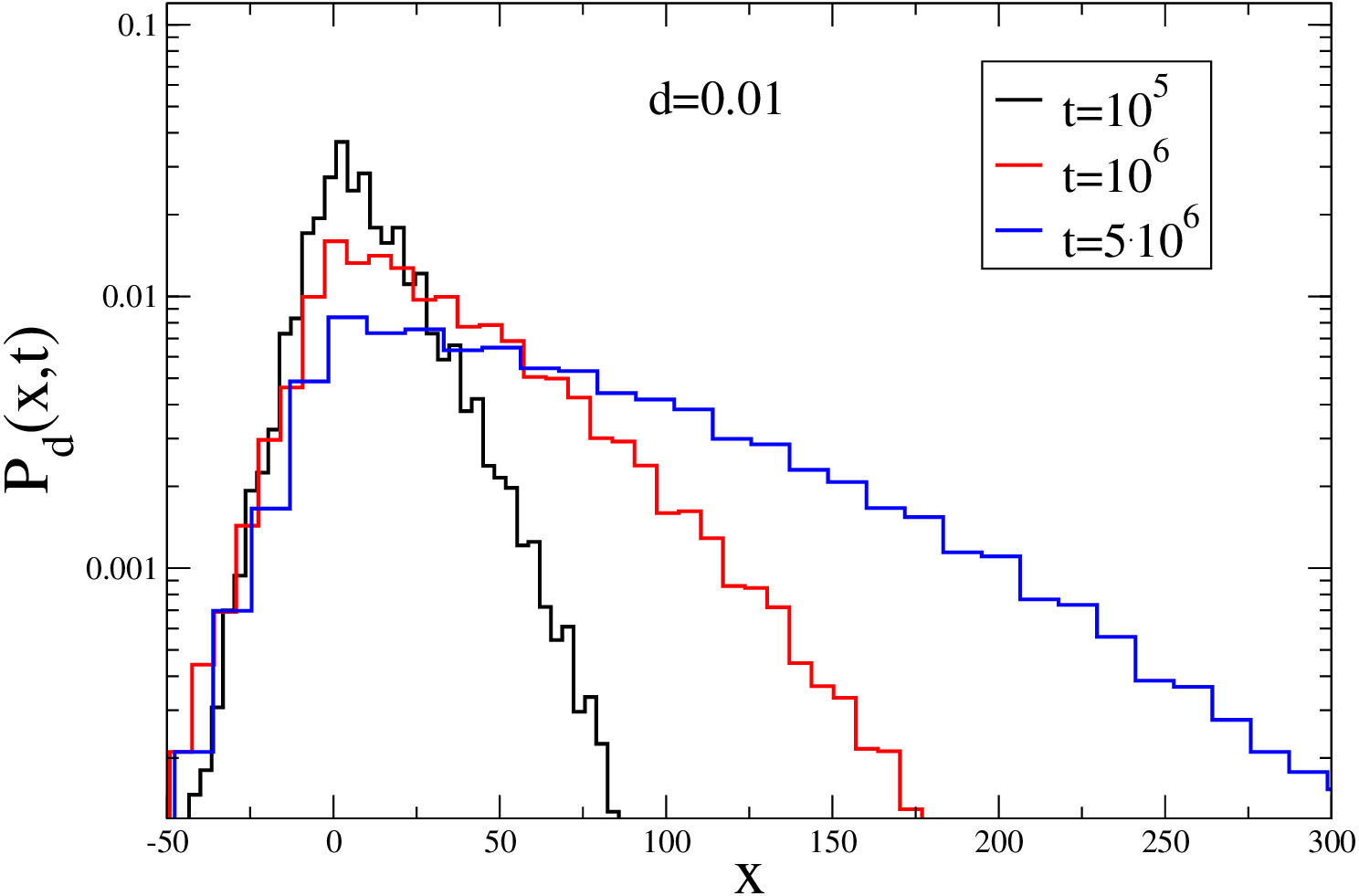}
\caption{$P_d(x,t)$ in the comb model with $L=\infty$ and $d=0.01$ at
different times. Notice that the mean value increases with time
mostly due to the spreading, while the most probable value
remains always close to zero.}
\label{pdf}
\end{figure}

\section{Conclusions and perspectives}

In order to evaluate the generality of the above results, let us
conclude by discussing another system.  Indeed, subdiffusion is
present in many different problems where geometrical constraints play
a central role. In this framework, a well studied phenomenon is the
so-called single-file diffusion.  Namely, we have $N$ Brownian rods on
a ring of length $L$ interacting with elastic collisions and coupled
with a thermal bath.  The equation of motion for the single particle
velocity between collisions is
\begin{equation}
m\dot{v}(t)=-\gamma v(t)+\eta(t),
\label{lang}
\end{equation}
where $m$ is the mass, $\gamma$ is the friction coefficient, and
$\eta$ is a white noise with variance
$\langle\eta(t)\eta(t')\rangle=2T\gamma\delta(t-t')$.  The combined
effect of collisions, noise and geometry (since the system is
one-dimensional the particles cannot overcome each other) produces a
non-trivial behaviour.  In the thermodynamic limit,
i.e. $L$,$N\to\infty$ with $N/L\to \rho$, a subdiffusive behaviour
occurs~\cite{HKK96}.

\begin{figure}[t!]
\centering
\includegraphics[scale=0.4,clip=true]{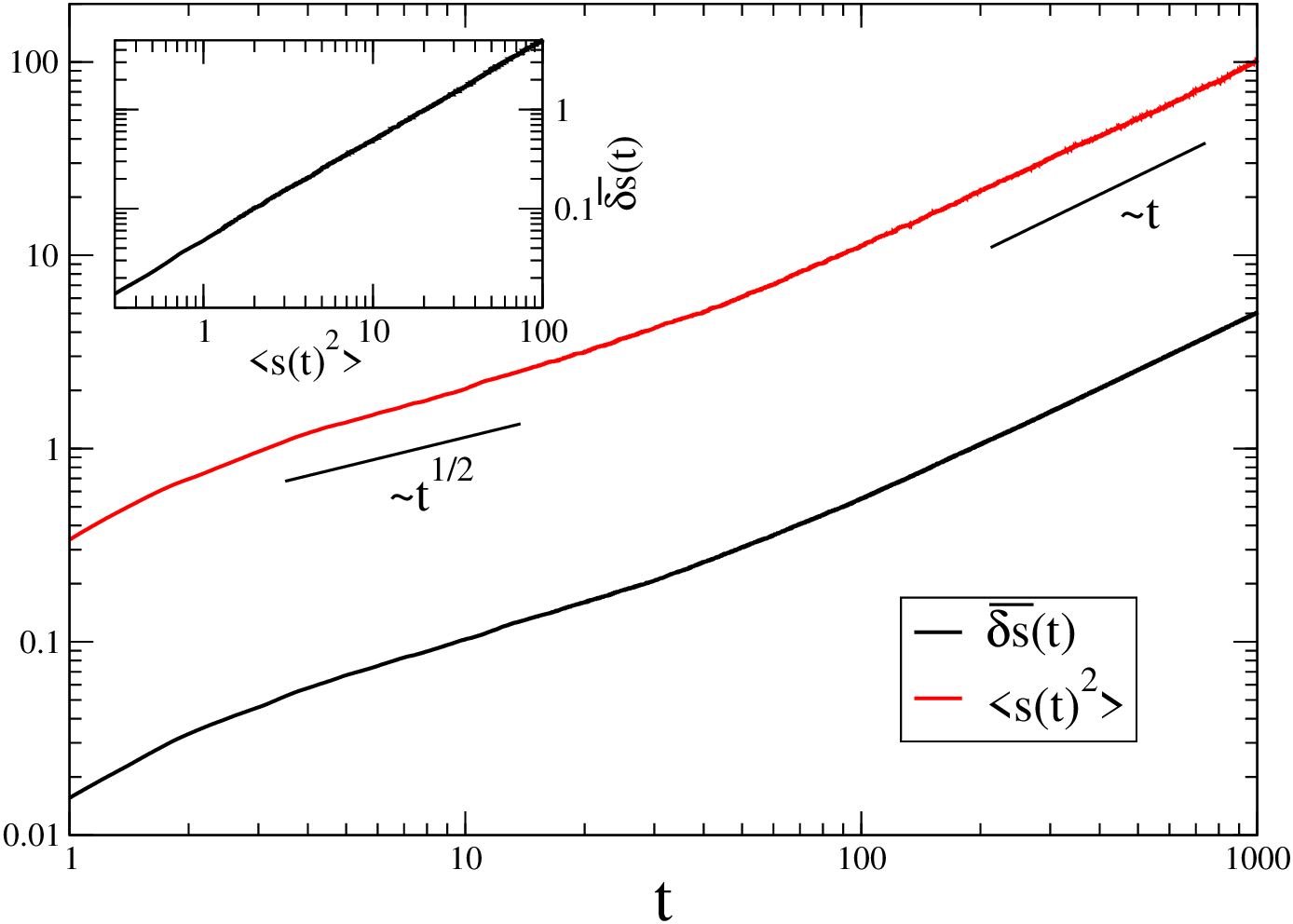}
\caption{$\langle s^2(t) \rangle$ and the response function
  $\overline{\delta s}(t)$ for the single-file model with parameters:
  $N=10$, $L=10$, $\sigma=0.1$, $m=1$, $\gamma=2$, $T=1$ and
  perturbation $F=0.1$. In the inset the parametric plot
  $\overline{\delta s}(t)$ \emph{vs} $\langle s^2(t) \rangle$ is
  shown.}
\label{single}
\end{figure}

Analogously to the comb model, the
case of $N$ and $L$ finite presents some interesting aspects.  In
order to avoid trivial results due to the periodic boundary conditions
on the ring, it is suitable to define the position of a tagged
particle as $s(t)=\int_{0}^{t}v(t')dt'$, where $v(t)$ is its
velocity. 
For the m.s.d. $\left< s^{2}(t)\right>$, averaged over the
thermalized initial conditions and over the noise, we find, after a
transient ballistic behaviour for short times, a dynamical crossover
between two different regimes:
\begin{equation}
\langle s^2(t) \rangle \simeq
\left\{\begin{array}{cc}
\frac{2(1-\sigma \rho)}{\rho
 }\sqrt{\frac{D}{\pi}}t^{1/2} &\phantom{mmm}  t < \tau^{*}(N)\nonumber
\\ 
 \frac{2D}{N} t &\phantom{mmm}  t >\tau^{*}(N),
\end{array}\right.
\end{equation}
where $\sigma$ is the length of the rods and $D$ is the diffusion
coefficient of the single Brownian particle~\cite{HKK96}.  Note that
the asymptotic behaviour is completely determined by the motion of the
center of mass, which is not affected by the collisions and simply
diffuses.  Moreover, as evident from numerical simulations,
$\tau^{*}\sim N^{2}$ and in the limit of infinite number of particles
the behaviour becomes subdiffusive, in perfect analogy with what
observed for the comb model, where the role of $L$ is here played by
$N$. The main difference
is that, in this case, the probability distribution is Gaussian in
both regimes.  As a consequence of the Gaussian nature of the problem,
applying a perturbation as a small force $F$ in Eq.~(\ref{lang}), one
finds that the Einstein relation is always
fulfilled~\cite{VPV08,LATBL10,BPRV08,PBV07}, also for finite $N$ and
$L$ (see Fig.~\ref{single}).
Strong violations of the Einstein
relation, can be obtained, in dense cases, 
when the collisions between the rods are inelastic so that
a homogeneous energy current crosses the system~\cite{VPV08}. \\

In this note we have considered systems with subdiffusive behaviour,
showing that the proportionality between response function and
correlation breaks down when ``non equilibrium'' conditions are
introduced.  In the comb model, non equilibrium effects are induced by
unbalanced transition probabilities driving the particle along the
backbone, while the single-file model is driven away from equilibrium
by inelastic collisions.  In the first case, the generalized FDR of
Eq.~(\ref{FDR}), developed in the framework of aging
systems~\cite{LCZ05}, can be explicitly written, providing the off
equilibrium corrections to the Einstein relation.  In the second case,
the transition rates are not known and another formalism must be
exploited~\cite{BPRV08}, which requires the
knowledge of the probability distribution for the relevant dynamical
variables of the model.  For instance, following the ideas
of~\cite{VPV08}, a distribution which couples the velocities of
neighbouring particles could be a reasonable guess.  Still, the
indentification of the relevant variables and their coupling in the
single-file and other granular systems is a central issue, requiring
further investigations.

\ack

We thank R. Burioni and A. Vezzani for interesting discussions on
random walks on graphs. We also thank F. Corberi and E. Lippiello for
a careful reading of the manuscript.  The work of GG, AS, DV and AP
is supported by the ``Granular-Chaos'' project, funded by Italian MIUR
under the grant number RBID08Z9JE.

\section*{References}
\bibliographystyle{unsrt}
\bibliography{paper.bib}

\end{document}